\documentclass[english]{jacow}
\usepackage[T1]{fontenc}
\usepackage[latin9]{inputenc}
\usepackage{textcomp}
\usepackage{amsmath}
\usepackage{graphicx}

\makeatletter

\providecommand{\tabularnewline}{\\}

\makeatother

\usepackage{babel}
\begin{document}
\title{Control algorithm tests using a virtual CW SRF cavity}
\author{Josu Jugo\thanks{corresponding author}, Ander Elejaga, University
of the Basque Country UPV/EHU, Leioa, Spain\\
Pablo Echevarria, Helmholtz Zentrum Berlin, Berlin, Germany}
\maketitle
\begin{abstract}
Superconducting cavities (SRF) are widely used in new generation particle
accelerators, increasing the requirements and specifications for new
designs. The LLRF control system, including the detuning control due
to mechanical perturbations, must fulfill more exigent specifications,
and its design have gained increasing relevance. The Helmhoz Zentrum
Berlin, among others, have been working in the development of simulation
and Hardware-in-the-loop tools to facilitate the test of control algorithms.
The main goal of this work is to use an existing cavity model in CW
mode, a Tesla cavity including a Saclay style piezo-tuner, and simulation
tools to compare and test different control strategies focused in
the detuning reduction, specially microphonics. The design process
consist of the use of pure simulation environment based on Matlab/Simulink,
where the mathematical model includes a cavity model, a LLRF control
system and detuning control strategies, considering the mentioned
actuator. Different control strategies are considered for the RF and
mechanical parts: perturbation reduction by PID based feedback loops,
adaptive feedforward algorithms, and active disturbance rejection
techniques (ADRC). The aim is the performance comparison of the different
algorithms with different perturbations, by using realistic cavity
models which include Lorenz force detuning, microphonics derived from
the cryogenic module and so forth. The simulation environment allows
the inclusion of other effects as the non-collocated control problem.
\end{abstract}

\section*{Introduction}

The use of simulation models and Hardware-in-the-loop (HIL) techniques
for developing virtual cavities is a relevant tool for the setting
up and debugging of cavities'' support systems, \cite{Branlard_et_al,HZB_1,HZB_2,Serrano,Branlard}.
This includes the analysis, design and test of LLRF control systems,
quench detection, resonance frequency control among others, resulting
in a reduction of the otherwise time consuming and costly operation
of real cavities.

On the other hand, simulation and HIL models are complementary. Simulation
models allow a versatile woking environment for rapid test and comparison
of different ideas and proposal \cite{Branlard_et_al}, while FPGA-based
virtual cavities allow the use of real hardware implementations, given
a more realistic picture of the system performance \cite{HZB_1,HZB_2,Serrano}
(including forward and reverse RF signal, quenching, field dependent
$Q_{0}$, Lorentz force detuning and microphonics) and the possibility
of a quasi automatic interchange of virtual and real cavities. However,
the implementation of the algorithms in a HIL system requires an extra
effort comparing with a simulation environment. As conclusion, in
a first phase, the use of simulated virtual cavities are useful for
an agile test and comparison of multiple ideas and algorithms and,
in a second phase, the most promising ones can be tested in a HIL
system, to validate the obtained results.

In this work, a simulated virtual TESLA cavity in CW mode \cite{Schilcher,neumann}
and LLRF system is used in order to test different control algorithm,
for the stabilization of the RF signal and for the frequency resonance
perturbation reduction. In particular, three control algorithms are
implemented: a standard PID control, an active disturbance rejection
control (ADRC) algorithm \cite{ADCR_SRF,Barnejee} and an adaptive
feedforward controller \cite{widrow}. Different combinations of such
control algorithms are used in the two control loops: the RF control
and the resonance frequency control loop. The cavity model considered
in this study as reference describes mainly a Tesla cavity from HZB.

Using different control scheme combinations, the effect of the interaction
between the two control loops can be also analyzed , since the relevance
of such interaction has also been reported in previous works. The
presented results show that this framework will be useful to study
different control strategies and problems, and encourage to perform
more detailed tests.

\section*{System description}

The simulations are based in the use of a 9-cell Tesla cavity model
implemented ins a MATLAB/Simulink environment, whose basic electric
parameters are shown in Table 1, being $Q_{o}$ the quality factor
of the cavity, $R/Q$ represent the efficiency of the acceleration
process which is dependent on the  cavity's geometry and $f_{RF}$
is the nominal resonant frequency of the cavity. Additionally, the
Table 2 shows the main mechanical modes of the cavity (20 modes in
total).

\begin{table}
\begin{centering}
\begin{tabular}{|c|c|c|c|}
\hline 
 & $Q_{o}$ & $R/Q\,(\Omega)$ & $f_{RF}\,GHz$\tabularnewline
\hline 
\hline 
Tesla cavity & $5.0\,10^{10}$ & $900$ & 1.3\tabularnewline
\hline 
\end{tabular}
\par\end{centering}
\caption{Main parameters of Tesla cavity}

\vspace*{-0.2cm}
\end{table}

\begin{table}
\begin{centering}
\begin{tabular}{|c|c|c|c|c|}
\hline 
 & \hspace*{-0.08cm}Mode 1\hspace*{-0.08cm} & \hspace*{-0.08cm}Mode 2\hspace*{-0.08cm} & \hspace*{-0.08cm}Mode 3\hspace*{-0.08cm} & \hspace*{-0.08cm}Mode 4\hspace*{-0.08cm}\tabularnewline
\hline 
\hline 
 $\omega_{n}$ & 262 & 589 & 1079 & 1216\tabularnewline
\hline 
damping $\delta$ & \hspace*{-0.08cm}0.0025\hspace*{-0.08cm} & \hspace*{-0.1cm}0.0045\hspace*{-0.1cm} & \hspace*{-0.1cm}0.0036\hspace*{-0.1cm} & \hspace*{-0.1cm}0.0047\hspace*{-0.1cm}\tabularnewline
\hline 
\hspace*{-0.25cm}%
\begin{tabular}{c}
\hspace*{-0.15cm}Lorentz coupling\hspace*{-0.15cm}\tabularnewline
(Hz/(MV/m)\textasciicircum 2)\tabularnewline
\end{tabular}\hspace*{-0.25cm} & 0.5 & 0.5 & 0.5 & 0.5\tabularnewline
\hline 
\hspace*{-0.3cm}%
\begin{tabular}{c}
Piezo coupling\tabularnewline
(Hz/V)\tabularnewline
\end{tabular}\hspace*{-0.3cm} & \hspace*{-0.08cm}-0.008\hspace*{-0.08cm} & \hspace*{-0.1cm}0.048\hspace*{-0.1cm} & \hspace*{-0.1cm}-0.052\hspace*{-0.1cm} & \hspace*{-0.1cm}0.038\hspace*{-0.1cm}\tabularnewline
\hline 
\end{tabular}
\par\end{centering}
\caption{Main mechanical modes of Tesla cavity }
\vspace*{-0.8cm}
\end{table}

\begin{figure}
\centering{}\includegraphics[scale=0.5]{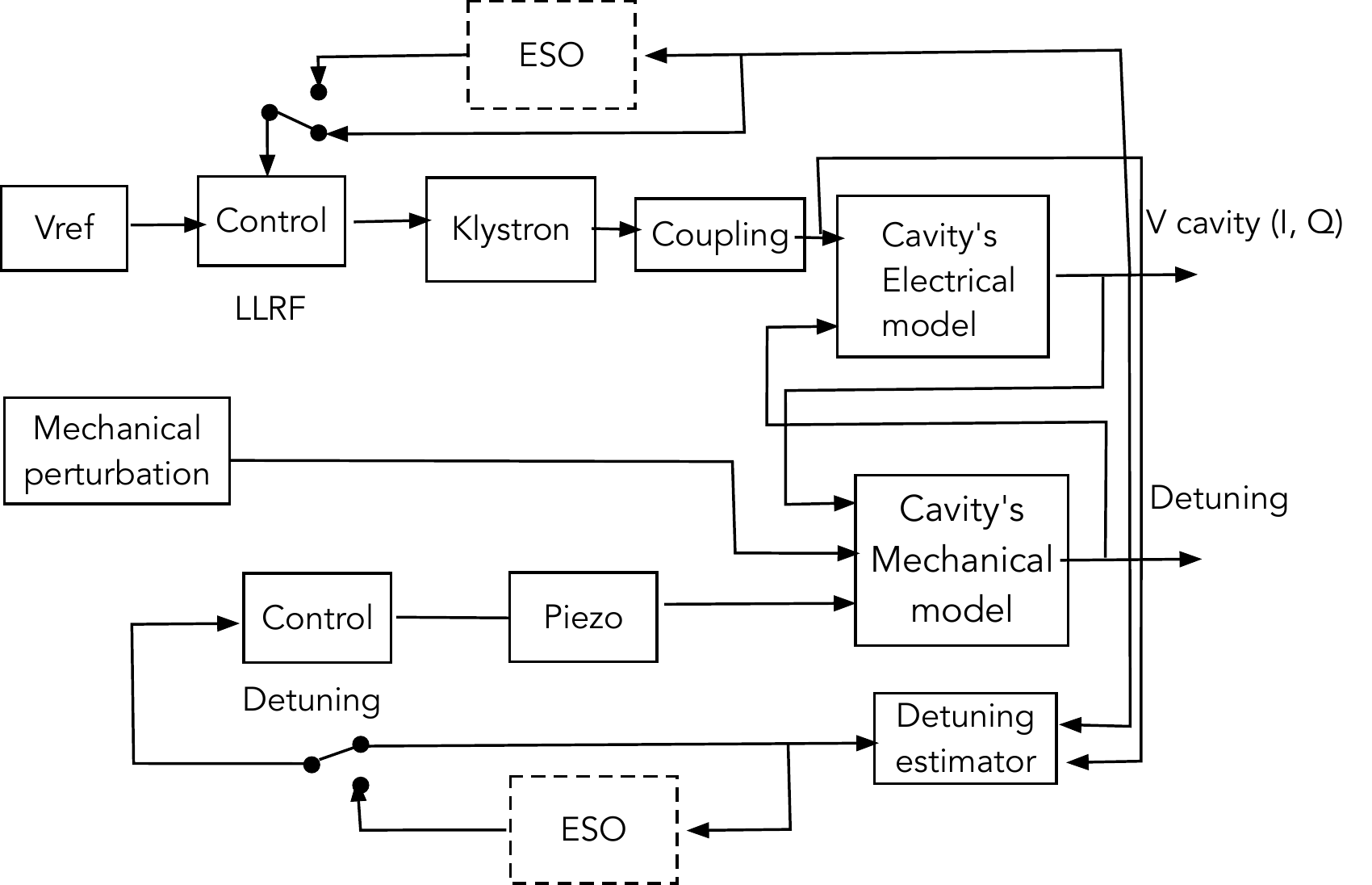}\caption{\label{fig:Simulation-scheme}Simplified simulation scheme, including
Klystron, coupling and cavity's electrical and mechanical model. The
observer blocks (ESO) are used for the ADRC controllers.}
\end{figure}

The scheme of the simulation system is depicted in Figure \ref{fig:Simulation-scheme}.
The system includes the electrical and mechanical model of the cavity,
the RF part (Klystron and coupling) and a piezo tuner, considering
a time delay and the mechanical effect of such tuner in the cavity.
The impact of the tuner delay is discussed in the Simulation results
Section. Additionally, the differences between the mechanical dynamics
of the cavity and the piezo allow the study of the non-collocated
control problem. The non-collocated control problem appears when the
sensor and the actuator are placed in different positions. In this
case, the cavity acts as the sensor, translating the mechanical perturbations
to the RF part. On the other hand, the piezo-tuner is positioned in
a particular place, which can lead to controllability issues. 

\section*{Implemented control algorithms}

The implemented system has two main control loops. The first one is
the principal LLRF system for the control of the RF signal in amplitude
and in phase, using an I/Q based approach. The second control loop
is introduced for reducing the mechanical perturbations by mean of
a piezo tuner.

\subsection*{LLRF control for the cavity field}

Two algorithms are considered here for the comparison tests: a standard
PI control and an ADRC algorithm.

\subsubsection*{PI control.}

The scheme of the discrete PI controller implemented is basic, described
by the next expression:
\begin{equation}
g(z\text{)}=K_{p}+\frac{K_{i}T_{s}}{z-1}\label{eq:pid}
\end{equation}
being $T_{s}=1\times10^{-5}s$ the sampling period. In this case,
the controller parameters are $K_{p}=2\times10^{-6}$ and $K_{i}=5\times10^{-3}$,
designed initially using the first order model of the cavity filling
and tuned by simulation. Two control loops are implemented for the
real and imaginary parts of the RF signal, allowing the amplitude
and phase control.

\subsubsection*{ADRC control.}

The ADCR controller is based on an Extended State Observer (ESO),
which takes the measured process's input and output, and estimates
the underlying noise-free trend in real time. The ESO estimates the
total disturbance acting on the system which is then fed back into
the control scheme and cancelled via the ADRC law. As a result of
this cancelation, the plant is reduced to its simplest form which
can be easily controlled via proportional means.

Based on \cite{ADRC_Geng}, an ADRC control algorithm has been developed
and applied to this particular cavity model. The MIMO cavity equation
can be expressed in matrix form as follows.

\begin{align}
\left[\begin{array}{c}
\dot{V_{cr}}\\
\dot{V_{ci}}
\end{array}\right] & =\left[\begin{array}{cc}
-\omega_{1/2} & -\Delta\omega\\
-\Delta\omega & -\omega_{1/2}
\end{array}\right]\left[\begin{array}{c}
V_{cr}\\
V_{ci}
\end{array}\right]\nonumber \\
 & +\frac{R_{L}\omega_{1/2}}{m}\left[\begin{array}{c}
I_{rfr}\\
I_{rfi}
\end{array}\right]+\frac{R_{L}\omega_{1/2}}{m}\left[\begin{array}{c}
I_{br}\\
I_{bi}
\end{array}\right]\label{eq:cavity}
\end{align}

\noindent where $V_{cr}$ and $V_{ci}$ are the real and imaginary
part of the cavity voltage, $\omega_{1/2}$ is the half-bandwidth
of the cavity, $R_{L}$ is the load resistance, m is the ratio of
the transformer and $\varDelta\omega$ is the detuning. Furthermore,
$I_{rfr}$, $I_{rfi}$, $I_{br}$ and $I_{bi}$ are the decomposition
into real and imaginary part of the effective driving intensities
for RF power and beam loading. Note that the aforementioned terms
can be defined in the following way.

$\boldsymbol{y}=\left[\begin{array}{c}
V_{cr}\\
V_{ci}
\end{array}\right]$, $\boldsymbol{A}=\left[\begin{array}{cc}
-\omega_{1/2} & -\Delta\omega\\
-\Delta\omega & -\omega_{1/2}
\end{array}\right]$, $B_{0}=\frac{R_{L}\omega_{1/2}}{m}$, $\boldsymbol{u}=\left[\begin{array}{c}
I_{rfr}\\
I_{rfi}
\end{array}\right]$, $\boldsymbol{d}=\left[\begin{array}{c}
I_{br}\\
I_{bi}
\end{array}\right]$

Being the beam loading viewed as a disturbance (\textbf{d}), the cavity
equation can be rewritten as

\begin{equation}
\dot{\boldsymbol{y}}=\boldsymbol{Ay}+B_{0}\boldsymbol{u}+B_{0}\boldsymbol{d}\label{eq:basic_adrc}
\end{equation}

If the system dynamics are considered as unknown perturbations, the
equation can be rewritten as follows.

\begin{equation}
\dot{\boldsymbol{y}}=\boldsymbol{f}+B_{0}\boldsymbol{u}\label{eq:basic_adrc1}
\end{equation}
where \textbf{f }is the general disturbance term that will be estimated
by the ESO, and further on actively cancelled by the ADRC control.
In this way, the two items to be estimated by the observer are $\hat{\boldsymbol{x}}_{1}=\hat{\boldsymbol{y}}$
and $\hat{\boldsymbol{x}}_{2}=\hat{\boldsymbol{f}}$. An observer
can be defined to estimate those two parameters based on the input
\textbf{u }and output \textbf{y }of the system\textbf{:}

\begin{equation}
\left[\begin{array}{c}
\hat{\dot{\boldsymbol{x}_{1}}}\\
\hat{\dot{\boldsymbol{x}_{2}}}
\end{array}\right]=\left[\begin{array}{cc}
0 & \boldsymbol{I}\\
0 & 0
\end{array}\right]\left[\begin{array}{c}
\boldsymbol{\hat{x}}_{1}\\
\boldsymbol{\hat{x}}_{2}
\end{array}\right]+B_{0}\left[\begin{array}{c}
\boldsymbol{I}\\
0
\end{array}\right]\boldsymbol{u}+\boldsymbol{L}(\boldsymbol{y-\hat{x}_{1}})\label{eq:adrc}
\end{equation}

\noindent where \textbf{L }is a parameter matrix that will determine
the poles of the observer and hence, its dynamics \cite{ADRC_Geng}.
Equation \ref{eq:adrc} can be rewritten in this particular case as

$\left[\begin{array}{c}
\dot{\hat{V}}_{cr}\\
\dot{\hat{V}}_{ci}\\
\dot{\hat{f}}_{r}\\
\dot{\hat{f}}_{i}
\end{array}\right]=\left[\begin{array}{cccc}
-l_{11} & -l_{12} & 1 & 0\\
-l_{21} & -l_{22} & 0 & 1\\
-l_{31} & -l_{32} & 0 & 0\\
-l_{41} & -l_{42} & 0 & 0
\end{array}\right]\left[\begin{array}{c}
\hat{V}_{cr}\\
\hat{V}_{ci}\\
\hat{f}_{r}\\
\hat{f}_{i}
\end{array}\right]+\left[\begin{array}{cccc}
l_{11} & l_{12} & B_{0} & 0\\
l_{21} & l_{22} & 0 & B_{0}\\
l_{31} & l_{32} & 0 & 0\\
l_{41} & l_{42} & 0 & 0
\end{array}\right]\left[\begin{array}{c}
V_{cr}\\
V_{ci}\\
I_{rfr}\\
I_{rfi}
\end{array}\right]$

The dynamics of the observer must be notably faster than the closed
loop behavior of the cavity, so that the estimation and rejection
of disturbances can be performed before they generate large perturbations
in the EM fields. For this particular system all the poles have been
placed at -3000 Hz and to do so, the \textbf{L }matrix parameters
have been determined as follows:

\begin{align*}
l_{12} & =l_{21}=l_{32}=l_{41}=0\\
l_{11} & =l_{22}=12000\pi\\
l_{31} & =l_{42}=36000000\pi^{2}
\end{align*}

Regarding the controller, a proportional control and disturbance rejection
can be implemented using the following control law

\begin{equation}
\boldsymbol{u}=\frac{\boldsymbol{K_{p}}(\boldsymbol{r-\hat{y}})-\boldsymbol{\hat{f}}}{B_{0}}\label{eq:adrc_control}
\end{equation}

\noindent being $\boldsymbol{K_{p}}$ a 2x2 gain matrix and \textbf{r
}the set point voltage of the cavity in real and imaginary form.

\subsection*{Control of mechanical perturbations}

Three different control algorithms are considered for the Lorenz force
detuning and microphonics reduction: a standard PI feedback loop ,
an ADRC controller and an adaptive feedforward

With respect to the mechanical disturbances suffered by the system,
two main sources have been considered. On the one hand, the cavity
voltage, set at 9 MV, generates a Lorenz Force detuning of about 600
Hz. On the other hand, to simulate the effect of external microphonics,
a white noise signal has been added to the mechanical model of the
cavity, generating a random detuning in the range from -5 Hz to 5
Hz, with a RMS value of 1.7 Hz. Furthermore, a pure sinusoidal signal
at $80Hz$ has been added.

\subsubsection*{PI control.}

A discrete PI controller (Eq. \ref{eq:pid}) has been implemented
as first choice. Similarly to the previous case, the controller has
been tuned by simulation and its control parameters are $K_{p}=0.55$
and $K_{i}=20$.

\subsubsection*{ADRC control.}

The model of the mechanical dynamics introduced by the Saclay II piezo
tuner used in this work has a relative order of one. This means that
a first order extended state observer (ESO) is enough to implement
the ADRC control \cite{ADCR_SRF,ADCR_SRF_1}. In this way, the parameters
to be observed are the detuning $\Delta\omega$ and the total disturbance
f. Following the process exposed in the section above, the observer
is defined by the next matrix equation.

\begin{equation}
\left[\begin{array}{c}
\hat{\dot{\Delta\omega}}\\
\hat{\dot{f}}
\end{array}\right]=\left[\begin{array}{cc}
-l_{1} & 1\\
-l_{2} & 0
\end{array}\right]\left[\begin{array}{c}
\hat{\Delta\omega}\\
\hat{f}
\end{array}\right]+B_{0}\left[\begin{array}{cc}
1 & l_{1}\\
0 & l_{2}
\end{array}\right]\left[\begin{array}{c}
V_{piezo}\\
\Delta\omega
\end{array}\right]\label{eq:cavity_adrc}
\end{equation}

\noindent where $\hat{\Delta\omega}$ and $\hat{f}$ are the observed
detuning and total disturbance, $V_{piezo}$ is the voltage entering
the piezo and $l_{1}$ and $l_{2}$ are the parameters that define
the dynamics of the observer. In this particular case, it is considered
that the piezo-tuner has a 114 $\mu s$ mechanical delay, and trying
to cancel every single disturbance would result in the destabilization
of the controller, due to the inability of the actuator to respond
in time to the fastest dynamics. In order to solve this issue, the
poles of the observer have been placed at a relatively low frequency
(50 Hz) to ensure that only the slowest disturbances are corrected.
This is an important limitation of the implemented design. In this
way, the L matrix parameters are defined as $l_{1}=628$ and $l_{2}=98696$
and $B_{0}=3.5714\times10^{-05}$.

\subsubsection*{Adaptive Feedforward.}

To cope with constant frequency microphonics, such as those provoked
by repetitive disturbances as vacuum pumps, an Adaptive Feedforward
(AFF) control \cite{widrow} has been tested. As it has been reported
in previous work, this kind of controllers notably suppress located
microphonics by computing and generating a sinusoidal control signal
with the appropriate amplitude and phase to cancel the perturbation.
In this manner, an AFF has been added to the existing PI and ADRC
controllers to try and suppress a 80 Hz microphonic. The particular
controller structure implemented, a filtered-x LMS algorithm, is shown
in the Figure 2.

\begin{figure}
\centering{}\includegraphics[scale=0.45]{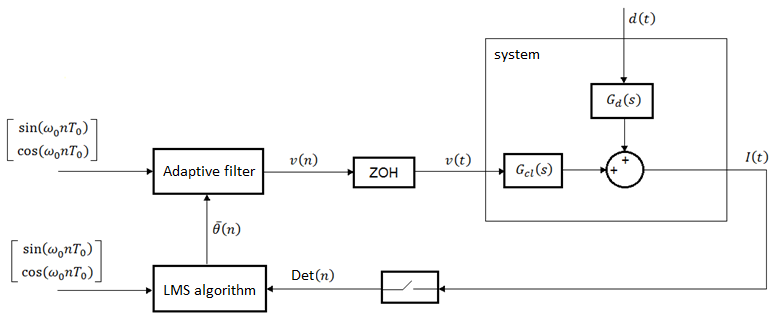}\caption{AFF control scheme}
\end{figure}

On the one hand, the LMS algorithm searches for the correct phase
and amplitude of the control signal by implementing the following
equations.

\begin{align}
\bar{\theta}(n) & =\left[A_{1}(n),A_{2}(n)\right]\nonumber \\
A_{1}(n) & =A_{1}(n-1)+\gamma Det(n)sin(\omega_{0}nT_{0})\nonumber \\
A_{2}(n) & =A_{2}(n-1)+\gamma Det(n)cos(\omega_{0}nT_{0})\label{eq:AFF}
\end{align}

\noindent being $Det(n)$ the detuning perturbation to reduce, $T_{0}$
is the AFF sampling rate and $\omega_{0}$ the frequency of the perturbation.
On the other hand, the adaptive filter generates a sine wave with
the parameters obtained from the LMS algorithm.

\begin{equation}
\upsilon(n)=\left[A_{1}(n),A_{2}(n)\right]\left[\begin{array}{c}
sin(\omega_{0}nT_{0})\\
cos(\omega_{0}nT_{0})
\end{array}\right]\label{ec:AFF1}
\end{equation}

\section*{Simulation results}

\begin{figure}
\begin{centering}
\includegraphics[scale=0.265]{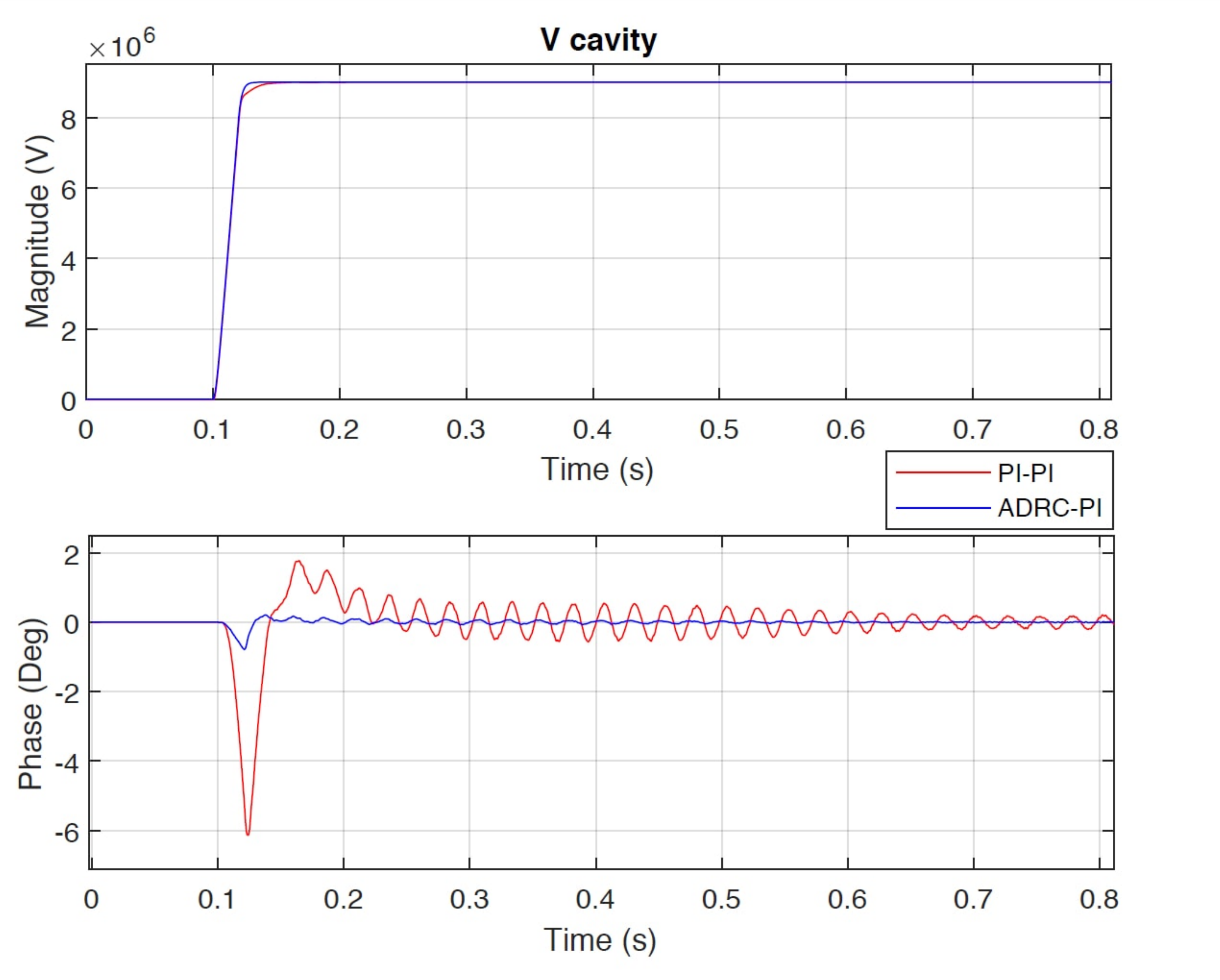}
\par\end{centering}
\caption{\label{fig:Time-response-of}Time response of the cavity filling}
\end{figure}

Different scenarios have been considered, combining the control algorithms
in the two loops, always in CW mode. Mainly, the next combinations
have been tested:
\begin{itemize}
\item Scenario 1: A basic PI in the 1st and 2nd loops
\item Scenario 2: An ADRC in the 1st loop and a PI controller in the 2nd
loop
\item Scenario 3: An ADRC in the 1st and 2nd loops
\item Scenario 4: A PI controller in the 1st loop and an ADRC controller
in the 2nd loop
\item Scenario 5: An ADCR controller in the 1st loop and a PI controller
and the adaptive feedforward controller in the 2nd loop
\end{itemize}
The figure \ref{fig:Time-response-of} shows the cavity filling, representing
the amplitude and phase behavior for the main two scenarios considered
for the 1st loop: a standard PI and the ADCR control algorithm. As
it is observed, the amplitude error is negligible in both cases, being
faster the stabilization of the signal in the ADRC case. Most evident
is the advantage of this control scheme observing the phase behavior,
being the stabilization time very short and final error very low (<0.01º).
In this figure, only scheme 1 and 2 are shown, since the effect of
the second control loop is negligible.

Figure 4 and 5 show the results obtained for the control of mechanical
perturbations. Four different scenario are included. The results of
the scheme 4 are not included since it has not been possible the obtaining
of reasonable results. This problem is discussed below.

\begin{figure}
\begin{centering}
\includegraphics[scale=0.6]{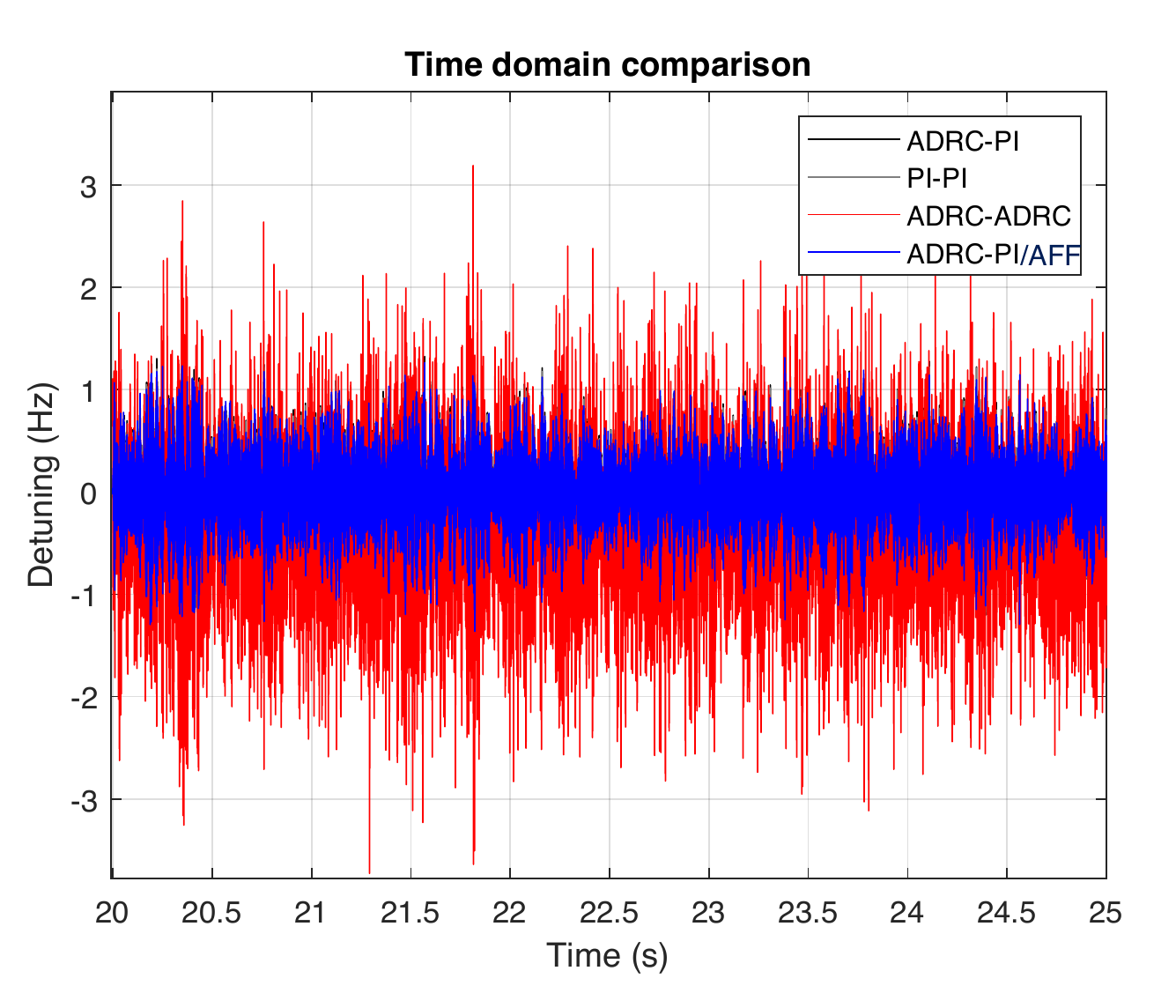}
\par\end{centering}
\caption{Microphonics control in the time domain. ADRC-PI, PI-PI and ADRC-PI/AFF
signals are overlapping.}
\end{figure}

As it is observed in those figures, the best results are obtained
with the basic PI controller in the tuner loop (Scheme 1, 2 or 5).
The PI controller reduces  the detuning, in the time domain, in a
factor 5 (comparing with the open-loop case), with an approximated
bandwidth of 350Hz. In figure 5, it is observed that the sinusoidal
perturbation is cancelled by mean of the AFF feedforward algorithm
(Scheme 5),

\begin{figure}
\begin{centering}
\includegraphics[scale=0.6]{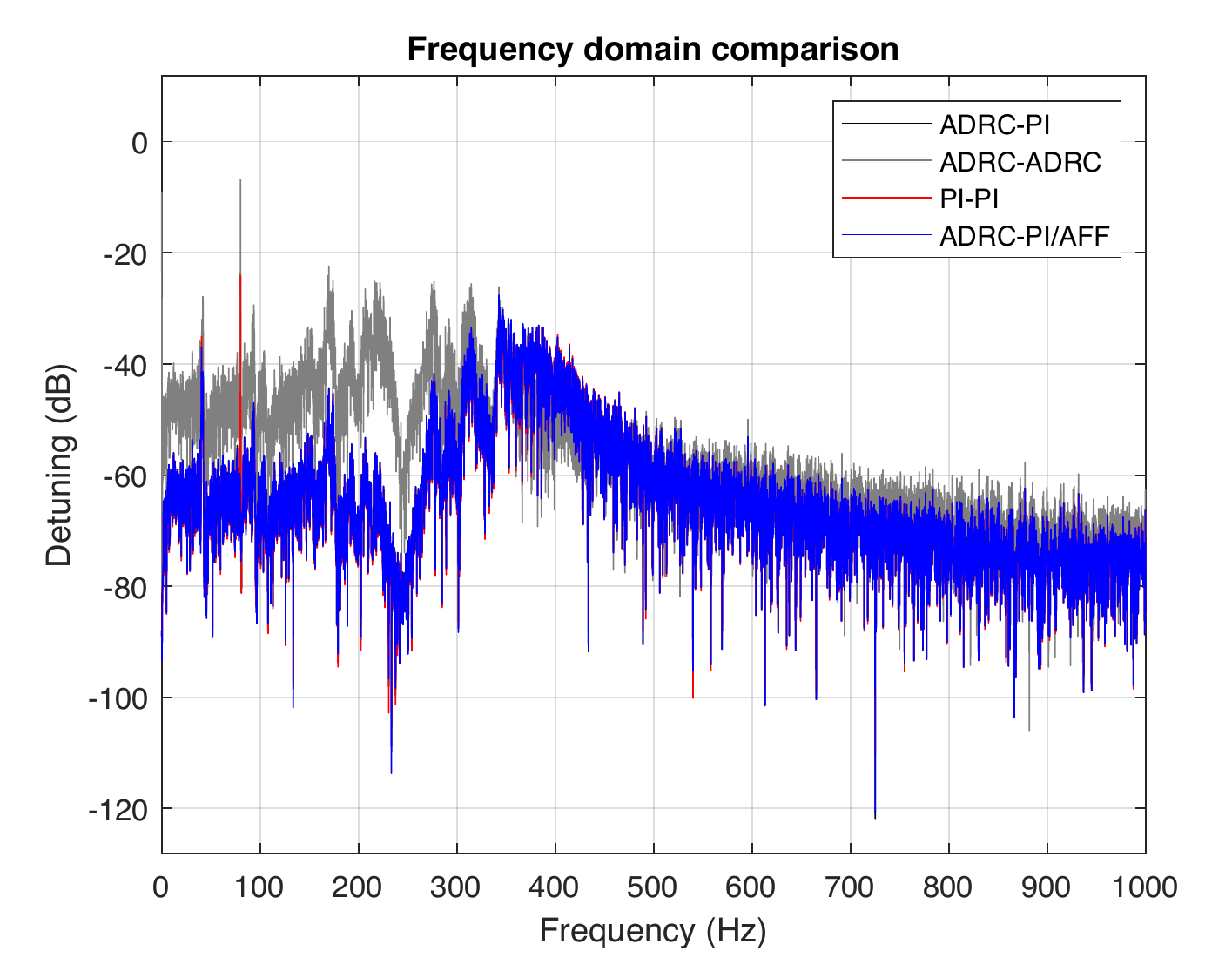}
\par\end{centering}
\caption{Microphonics control in the frequency domain}
\end{figure}

\begin{figure}
\begin{centering}
\includegraphics[scale=0.6]{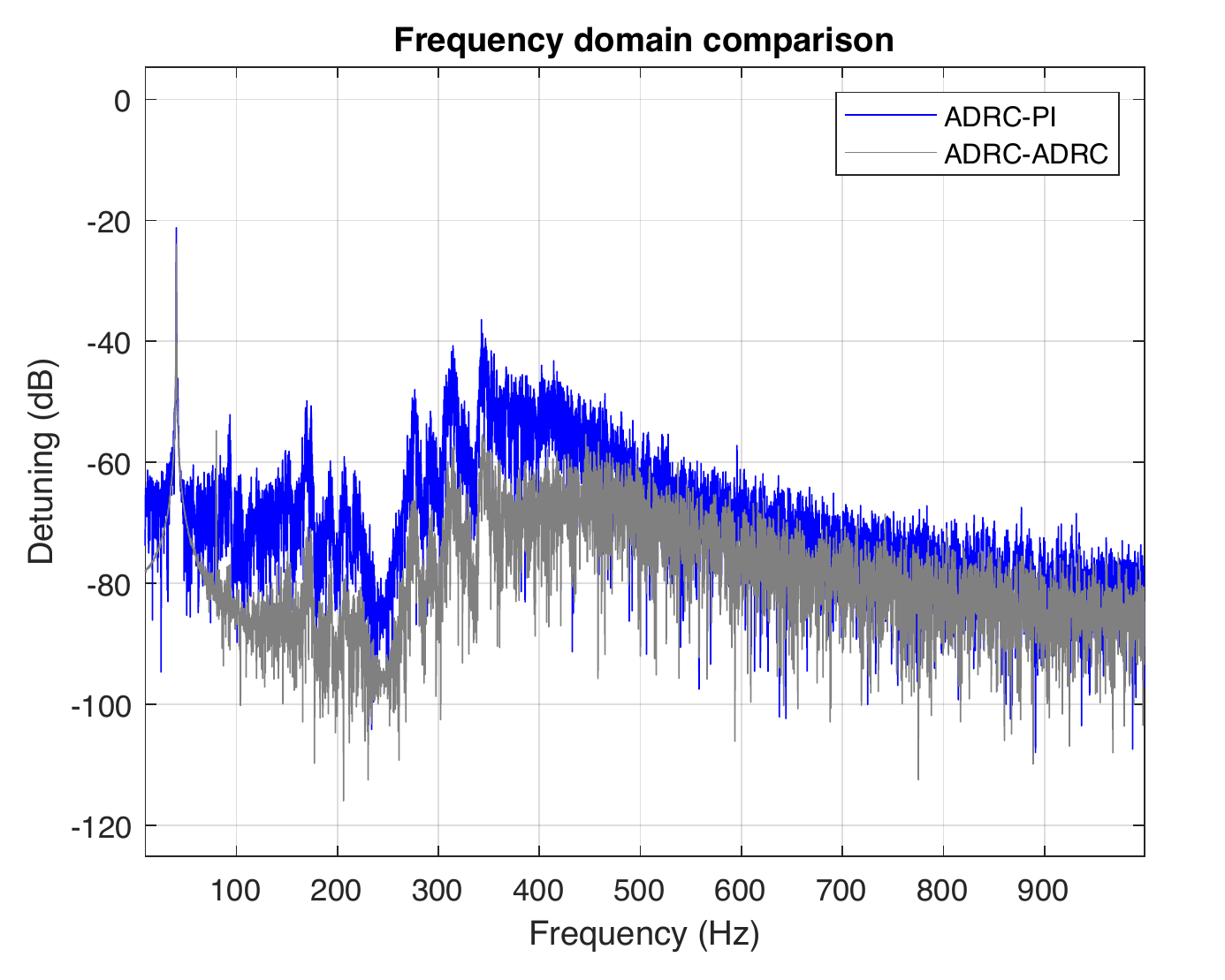}
\par\end{centering}
\caption{\label{fig:Microphonics-control-in}Microphonics control in the frequency
domain without tuner time delay}
\end{figure}

The ADRC algorithm has reported very good results comparing with PID
algorithms in the literature. This is the case for the first control
loop. However, in the control loop for the detuning reduction, the
results are not so good. The problem is the tuning of the parameters
of the ADRC controller (observer and control gains) for stabilizing
the loop, due to the delay introduced by the tuner. This delay has
a great impact in the ADRC performance, limiting its effect. As it
is shown in Figure \ref{fig:Microphonics-control-in}, the ADRC controller
gives very good results, reducing the delay effect.

On the other hand, the delay and the interaction of the ADRC controllers
in both loops (Scenario 4) cause the instability of the system.

Finally, the non-collocated control problem is considered supposing
different dynamics for the mechanical perturbations in the cavity
and the actuator's effect, the piezo-tuner. The control schemes considered
have been able to work in the particular configuration studied. Nonetheless,
more analysis are needed, since the controllability or observability
loss due to the non-collocated control problem can lead to malfunction
controllers.

\section*{Conclusions}

The use of virtual cavities facilitates the improvement of its support
systems and, in particular following the model design approach, the
test of different control algorithms for stabilizing the RF signals
and the resonance frequency against perturbations. Taking into account
the results presented in this work, the ADRC control strategy, which
has been successfully tested in previous works, and its combination
with other control techniques shows very good results in the simulation
test, which fuels further studies in the future. The preliminary simulation
results show that delay has an important impact in the performance
of the microphonics reduction.

Consequently, the reduction of the delay effect in the system performance,
implementing adequate control schemes \cite{ADCR_delay1,ADCR_delay},
is a problem to consider in future works. Other problem to consider
is the non-collocated control problem, that is, performance considerations
taking into account that perturbation sources and correction actions
are located physically at different points.

In any case, those good (and future) results, should be validated
in a second phase, using the FPGA based HIL system, since provides
a more realistic simulator and will meet the system\textquoteright s
dynamics in time and will allow testing the controller\textquoteright s
hardware implementation as well as its in-time performance. This procedure
will lead to a better performance of the Tesla cavity under analysis.

However, as a more general consideration, the use of open source cavity
models as reference for testing of control algorithms will give information
not only valid for a particular system, but general insights for selecting
the most promising control algorithms to all the community.

\section*{Acknowledgement}

The authors are very grateful for the partial support of this work
by the projects DPI2017-82373-R ( (Spanish Ministry of the Economy,.
Industry and Competitiveness) and PIT30 (ELKARTEK, Program of the
Basque Government).

\end{document}